\documentclass{article}
\usepackage{frascatiphys_SP}
\usepackage{amssymb}
\usepackage{epsfig}
\usepackage{eucal}
\usepackage{amsmath}

\begin{document}
\title{ Nucleon form factors measurements via the radiative 
 return at~$\boldsymbol{ B}$-meson factories
\thanks{\hspace{0.2cm} Supported in part by 
EC 5th Framework Programme under contract 
HPRN-CT-2002-00311 (EURIDICE network).}}

\author{
E.~Nowak 
 \\
\em Institute of Physics, University of Silesia, Katowice, POLAND.}

\maketitle
\baselineskip=11.6pt

\baselineskip=11.0pt
\newcommand{\ta}[1]{#1\hspace{-.42em}/\hspace{-.07em}} 
\newcommand{\taa}[1]{#1\hspace{-.58em}/\hspace{-.09em}} 
\newcommand{\beq}{\begin{equation}} 
\newcommand{\eeq}{\end{equation}}   
\newcommand{\bea}{\begin{eqnarray}}
\newcommand{\eea}{\end{eqnarray}}
\newcommand{\non}{\nonumber}
\newcommand{\eq}[1]{eq.(\ref{#1})}         
\newcommand{\Eq}[1]{Eq.(\ref{#1})}
\newcommand{\e}{\varepsilon} 
\newcommand{\be}{\beta} 
\phantom{}
\vspace{+0.1cm}
\section{Introduction}

The present-day experimental situation concerning nucleon form factors
 in the space-like region shows a substantial 
 discrepancy between measurements via
 the Rosenbluth method and the recoil polarisation technique 
 (see Ref. \cite{arrington} for a review). 
 It was shown recently \cite{chen}, that the difference
 can be partly explained by two-photon mediated processes, that were 
 not taken into account in the original analysis.
 More information about nucleon form factors in the time-like region
  will not only improve poor experimental knowledge there, 
 but it will also shed light on the situation in the space-like region.
 The radiative return method \cite{Binner} is a powerful tool to provide that
 information using data of $B$- meson factories \cite{nukleony}, as we shall
 advocate also here.

\section{The radiative return and nucleon form factors measurements}

 To profit fully from the radiative return method, a Monte Carlo event
 generator is needed. For that purpose an upgraded version of 
 PHOKHARA \cite{Rodrigo:2001kf}
 (PHOKHARA 4.0) was developed \cite{nukleony}.
 It allows for a simulation of the reaction 
 $e^+e^-\to N\bar{N}\gamma (\gamma)$, where
 $N\bar{N}$ is a nucleon-antinucleon pair. It includes 
 initial state radiation (ISR) at next-to-leading order (NLO). 
 Basing on \cite{Czyz:}, we expect that the leading order (LO) final 
 state radiation (FSR) is negligible at $B$-factories, but
 NLO FSR (not included yet in the program)
  will be important for a measurement aiming for
 a few percent accuracy.

\subsection{The nucleon current}

The matrix element of the electromagnetic nucleon current
 is given by 
\begin{equation}
 J_\mu =  - i e \cdot \bar u(q_2)\left(F_1^N(Q^2)\gamma_\mu
 - \frac{F_2^N(Q^2)}{4 m_N} \left[\gamma_\mu,\taa Q \ \right]
 \right) v(q_1)~,
\label{prad}
\end{equation}
where $F_1$ and $F_2$ are the Dirac and Pauli form factors and $m_N$ 
is the nucleon mass. The antinucleon and nucleon momenta are denoted
 by $q_1$ and $q_2$ respectively, and $Q=q_1+q_2$. They are related 
to the magnetic and electric Sachs form factors by 
\begin{equation}
G_M^N = F_1^N + F_2^N~, \qquad  G_E^N = F_1^N + \tau F_2^N~,
 \quad {\rm with}\ \  \tau =Q^2/4m^2_N \ .
\nonumber
\end{equation}
 The parametrization of the form factors used in PHOKHARA follows from
 \cite{iach1,iach2} and is 
 in agreement \cite{nukleony} with the ratio of the form factors
  measured with the recoil polarisation method \cite{JLab1}.
\begin{figure}[t]
 \vspace{6.0cm}
\includegraphics{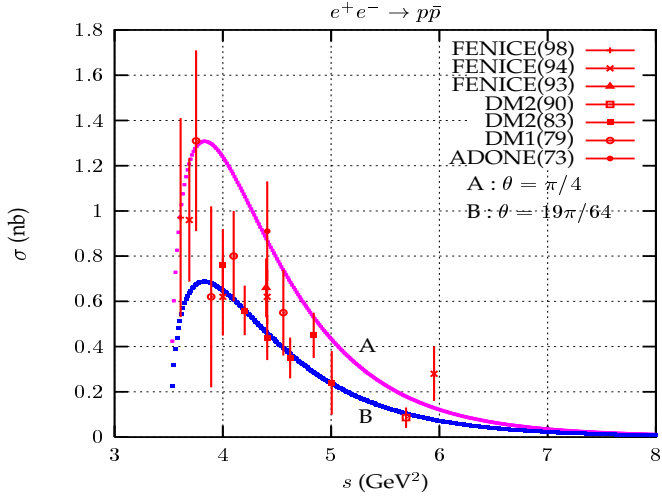}
 \vspace{-1.6cm}
\caption{\it Comparison of the measured 
\cite{antonelli}
$e^+e^- \to p\bar p$ cross section with the model from 
Ref.~\cite{iach1}. Predictions are given
for two different values ($\pi/4$ - curve A  and $19\pi/64$ - curve B)
 of the parameter $\theta$.
    \label{eepp} }
\end{figure}
 Available experimental data in the time-like region 
  consist only of total cross section measurements,
 and give practically no information about the form factors.
 Predictions for $\sigma(e^+e^-\to p\bar{p})$,
 $\sigma(e^+e^-\to n\bar{n})$
 and  $\sigma(p\bar{p}\to e^+e^-)$, obtained with the form factors
 used in PHOKHARA, are
 in good agreement with the data, 
 as shown  in Fig. \ref{eepp}
 for the reaction $e^+e^-\to p\bar{p}$.
 Other cross sections can be found in Ref. \cite{nukleony}.

\subsection{The method for the measurement of the nucleon form factors}

\begin{figure}[t]
 \vspace{6.0cm}
\includegraphics{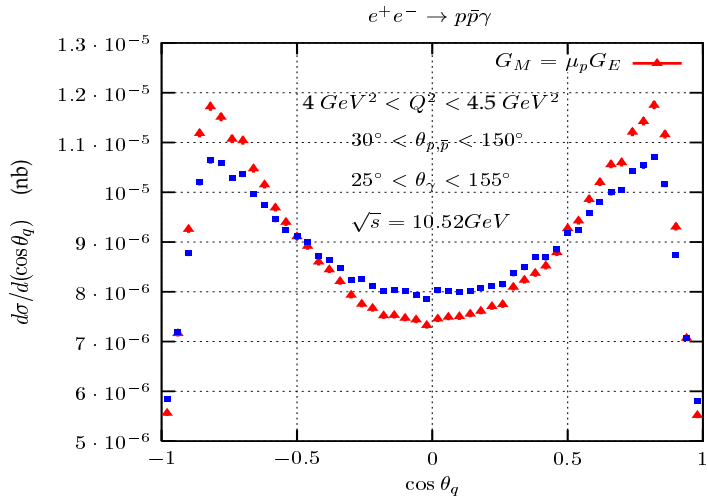}
 \vspace{-1.6cm}
 \caption{\it
      Angular distribution in the polar angle of vector
${\bf q}= ({\bf q}_2-{\bf q}_1)/2$ in~the CMS of the $e^+e^-$ pair.
    \label{me} }
\end{figure}
\begin{figure}[t]
 \vspace{6.0cm}
\includegraphics{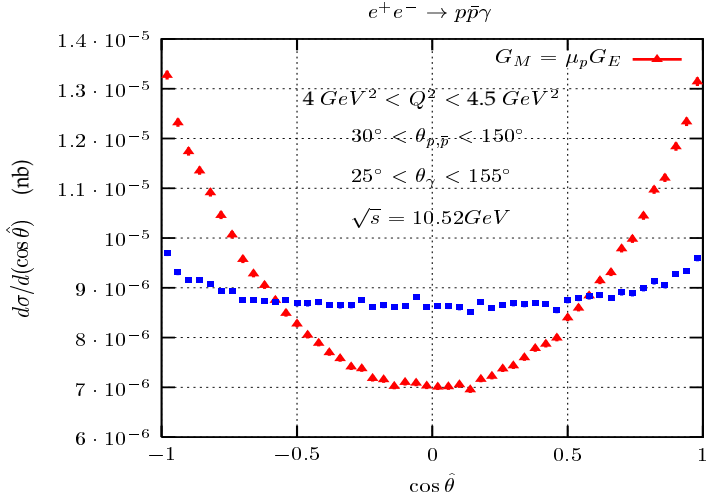}
 \vspace{-1.6cm}
 \caption{\it
    Angular distribution in the polar angle of vector
${\bf q}= ({\bf q}_2-{\bf q}_1)/2$  
  in the ${\bf Q}$-rest frame (${\bf q}={\bf q}_2$ in this frame).
    \label{mQ} }
\end{figure}

 The idea of the nucleon form factors measurements in the time-like 
 region via the radiative return is based on studies of angular distributions.
 The hadronic tensor of the process $e^+e^-\to N\bar{N}\gamma $
 depends only on $|G_M^N|^2$ and $|G_E^N|^2$ for 
 non-polarized nucleons \cite{nukleony}
 and thus
 it is not possible to measure the relative phase between $G_M^N$ and $G_E^N$,
 without measuring the nucleon polarisation.
 Distributions in the the ${\bf q}= ({\bf q}_2-{\bf q}_1)/2$ polar angle, for
 unpolarized nucleons, are shown in Figs. \ref{me} and \ref{mQ}. 
To show how sensitive  are the angular distributions to the form 
factors ratio, two predictions are presented:
 differential cross sections obtained for the model described
 above, and differential cross section with the assumption that
 $G^p_M=\mu_p G^p_E$ and the constraint
 that the $\sigma(e^+e^-\to p\bar{p})$ remains unchanged.
 The predicted number of events, for BaBar energy and 
 4~GeV$^2<$~Q$^2< $~4.5 GeV$^2$, is about 3500 with an accumulated
  luminosity of 200 fb$^{-1}$.
  It means, that
 a two parameter fit ($|G_M^N|$ and $|G_E^N|$) to the experimental 
 angular distributions,
 preferably in the ${\bf Q}$-rest frame
 (compare Fig. \ref{me} and Fig. \ref{mQ}), with relatively
 small Q$^2$ spacing,  
 is possible, and it will not be limited statistically.
\section{Summary}
 
 The Monte Carlo simulations with PHOKHARA 4.0 
 show that it is possible to measure separately
 the electric and magnetic nucleon form factors in time-like region 
 at $B$-meson factories by studying nucleon angular distributions 
 of events with emission of photons. 
 The radiative return is well suited for this measurement over 
 a wide kinematic range. 

\section{Acknowledgements}
The author would like to thank H.~Czy\.z and G.~Rodrigo for careful reading the manuscript.


%
\end{document}